\documentclass[preprint,aps,showpacs]{revtex4}

\input epsf

\def\Journal#1#2#3#4{{#1} {\bf #2}, #3 (#4)}

\def\NPB{{Nucl. Phys.} B}
\def\NPA{{Nucl. Phys.} A}
\def\PLB{{Phys. Lett.}  B}
\def\PRL{Phys. Rev. Lett.}
\def\PRD{{Phys. Rev.} D}
\def\PRC{{Phys. Rev.} C}
\def\ZPC{{Z. Phys.} C}

\def\JCP{J. Comp. Phys.}

\def\JPG{J. Phys. G}

\begin{document}

\title{Elliptic flow of gluon matter in \\
ultrarelativistic heavy-ion collisions}

\author{Zhe Xu \footnote{xu@th.physik.uni-frankfurt.de}
and Carsten Greiner}
\affiliation{Institut f\"ur Theoretische Physik, 
Goethe-Universit\"at Frankfurt, Max-von-Laue-Strasse 1, 
D-60438 Frankfurt am Main, Germany}

\date{\today}

\begin{abstract}
Employing a perturbative QCD based parton cascade we calculate
the elliptic flow $v_2$ and its transverse momentum dependence $v_2(p_T)$
for the gluon matter created in Au+Au collisions at $\sqrt{s_{NN}}=200$ GeV.
To make comparisons with the experimental data at the BNL Relativistic
Heavy Ion Collider (RHIC), parton-hadron duality is assumed.
We find that whereas the integrated $v_2$ matches the experimental data,
the gluon (or pion) $v_2(p_T)$ is about 20$-$50$\%$ smaller than the
experimental data. Hadronization via gluon fragmentation and quark 
recombination seems to be the key to explaining the necessary jump of
$v_2(p_T)$ from the partonic to the hadronic phase. We also show that
the elliptic flow values moderately depend on the chosen freezeout
condition, which will thus constrain the shear viscosity to the entropy
density ratio of the quark gluon plasma created at RHIC.
\end{abstract}


\maketitle

\section{Introduction}
\label{intro}
Elliptic flow \cite{O92,VZ96} is the key observable 
characterizing the collectivity \cite{VPS08} of the quark gluon plasma 
(QGP) that is potentially produced in ultrarelativistic heavy-ion 
collisions at the BNL Relativistic Heavy Ion Collider (RHIC) or/and 
at the CERN Large Hadron Collider (LHC). A fast transformation from the 
initial spatial anisotropy to the momentum anisotropy will indicate 
sufficiently strong interactions among quarks and gluons, which also 
provides the reason for a fast thermalization to a QGP with a small 
shear viscosity. The good agreements of the elliptic flow $v_2$ data at 
RHIC \cite{PHEN04,STAR04-1} with the results from calculations employing 
ideal hydrodynamics \cite{H01} suggest that the shear viscosity of the 
QGP created at RHIC is small \cite{L07}. To find its lower and upper bound 
is an important but difficult task, because the elliptic flow is built up 
not only during the evolution of the QGP but also initially during the 
thermalization and finally, though possibly marginal, during the hadronic 
cascade before particles decouple kinetically. Moreover, the lack of 
understanding of the hadronization leads to an uncertainty in converting 
the partonic elliptic flow to the hadronic one that is measured. All
these contributions to the final hadronic elliptic flow can, in principle,
be investigated in a transport calculation including a partonic and 
a hadronic cascade and a hadronization mechanism \cite{AMPT05}. 
Such studies will help to determine the true shear viscosity in the QGP phase.

Recently, employing the perturbative QCD (pQCD) based parton cascade
Boltzmann Approach of MultiParton Scatterings (BAMPS) \cite{XG05,XG07,EXG08},
we have calculated the elliptic flow $v_2$ of the gluon matter produced
in Au+Au collisions at the RHIC energy \cite{XGS08}.
To compare our results with the experimental data the gluonic $v_2$ was 
converted to the hadronic one using the parton-hadron duality picture
at a freezeout energy density of $e_c=1$ $\rm{GeV\ fm}^{-3}$. Hence
$v_2^{\rm{pion}}(p_T) = v_2^{\rm{gluon}}(p_T)$ at transverse momenta $p_T$.
Even though the hadronization was described in such
a simple manner, it was found that when employing a QCD coupling constant of
$\alpha_s=0.6$, the calculated $v_2$ matches the experimental
data \cite{star1,phobos1}. With $\alpha_s=0.3$ the results are about
$20\%$ smaller than the data. Hence, pQCD interactions can explain the
large $v_2$ buildup at RHIC. In addition, in Refs. \cite{XGS08} 
and \cite{XG08} 
the shear viscosity to the entropy density ratio $\eta/s$ from the transport
calculations was extracted and it was found that $\eta/s$ 
is between $0.15$ for $\alpha_s=0.3$ and $0.08$ for $\alpha_s=0.6$.
These findings are similar to those obtained from viscous hydrodynamical
calculations \cite{RR07}.

This article provides details on our recent results on the elliptic flow
from parton cascade calculations. In Sec.\ \ref{sec1} setups for numerical 
calculations are given. Section \ref{sec2} shows the results for
the transverse momentum dependence of the elliptic flow, $v_2(p_T)$, 
the transverse momentum spectra, the mean transverse momentum,
and the final transverse energy per rapidity at midrapidity. The
results from the BAMPS calculations are compared with the experimental data
at RHIC assuming parton-hadron duality. Further possible improvements
as well as a discussion about possible hadronization scenarios via 
gluon fragmentation and quark recombination are given. 
In Sec.\ \ref{sec3} the dependence of the elliptic flow on the chosen 
freezeout energy density is presented, which indicates the uncertainty 
of the $\eta/s$ ratio extracted from the BAMPS calculations. 
Finally we summarize in Sec.\ \ref{sum}.  In Appendix \ref{app1},
the calculation of the number of participating nucleons and 
the determination of centrality classes are given.

\section{BAMPS Setups}
\label{sec1}
BAMPS solves the Boltzmann equation for on-shell gluons with pQCD 
interactions, which include elastic scatterings and bremsstrahlung and 
its backreaction. The cross section of pQCD elastic scatterings is given 
by \cite{XG05,B93,W96}
\begin{equation}
\label{cs22}
\frac{d\sigma^{gg\to gg}}{d{\bf q}_{\perp}^2}=
\frac{9 \pi \alpha_s^2}{({\bf q}_{\perp}^2+m_D^2)^2}\,,
\end{equation}
where ${\bf q}_{\perp}$ denotes the perpendicular component of the momentum
transfer in the center-of-mass frame of the elastic collision. 
The interactions are screened by a Debye mass \cite{XG05,B93,W96},
\begin{equation}
\label{md}
m_D^2({\bf x},t)=\pi d_G \,\alpha_s N_c 
\int \frac{d^3p}{(2\pi)^3} \frac{f({\bf x},t,{\bf p})}{p}\,,
\end{equation}
where $d_G=16$ is the gluon degeneracy factor for $N_c=3$. $m_D$ is 
calculated locally using the gluon density function $f({\bf x},t,{\bf p})$
obtained from the BAMPS simulation. The integrated cross section is 
\begin{equation}
\label{tcs22}
\sigma^{gg\to gg}=\frac{1}{2}\int_0^{s/4} d{\bf q}_{\perp}^2\,
\frac{d\sigma^{gg\to gg}}{d{\bf q}_{\perp}^2}=
\frac{9}{2} \frac{\pi \alpha_s^2}{m_D^2 (1+4m_D^2/s)}\,,
\end{equation}
and $s$ is the invariant mass of the collision. The factor $1/2$ before the 
integral indicates that the outgoing particles in the collision are identical.

The effective matrix element of pQCD inspired bremsstrahlung
$gg\leftrightarrow ggg$ is taken in a Gunion-Bertsch form \cite{B93,W96,GB82},
\begin{equation}
\label{m23}
| {\cal M}_{gg \to ggg} |^2 =\frac{9 g^4}{2}
\frac{s^2}{({\bf q}_{\perp}^2+m_D^2)^2}\,
 \frac{12 g^2 {\bf q}_{\perp}^2}
{{\bf k}_{\perp}^2 [({\bf k}_{\perp}-{\bf q}_{\perp})^2+m_D^2]}\,
\Theta(k_{\perp}\Lambda_g-\cosh y)\,,
\end{equation}
where $g^2=4\pi\alpha_s$. ${\bf k}_{\perp}$ and $y$ denote the perpendicular
component of the radiated gluon momentum and its rapidity in the 
center-of-mass frame of the collision, respectively. 
$k_{\perp}=|{\bf k}_{\perp}|$, and $\Lambda_g$ is the
gluon mean free path, which is calculated self-consistently \cite{XG05}.
The suppression of the bremsstrahlung due to the Landau-Pomeranchuk-Migdal 
effect is effectively taken into account within the Bethe-Heitler regime using 
the step function in Eq. (\ref{m23}). Gluon radiations and absorptions
are only allowed if the formation time of the process, typically 
$\tau=\cosh y/k_{\perp}$, is shorter than the mean free path of the
radiated or absorbed gluon. Both the Debye screening mass and the gluon
mean free path act as an infrared regulator.

The total cross section for bremsstrahlung is obtained by the integral
\begin{equation}
\label{cs23}
\sigma_{gg \to ggg} = \frac{1}{2s} \frac{1}{3!} \int 
d\Gamma^{'}_1 d\Gamma^{'}_2 d\Gamma^{'}_3
| {\cal M}_{gg \to ggg} |^2 (2\pi)^4
\delta^{(4)}(p_1+p_2-p^{'}_1-p^{'}_2-p^{'}_3)\,,
\end{equation}
where $d\Gamma^{'}_i \equiv d^3 p^{'}_i/(2\pi)^3 /(2 E^{'}_i)$. 
$p_1$ and $p_2$ denote the four-momenta of the incoming gluons, and
$p^{'}_i$, $i=1$, $2$, $3$ denote the four-momenta of the outgoing gluons.
For the backreactions we define a ``cross section'' by \cite{XG05}
\begin{equation}
\label{cs32}
I_{ggg \to gg} = \frac{1}{2} \int d\Gamma^{'}_1 d\Gamma^{'}_2
| {\cal M}_{ggg \to gg} |^2 (2\pi)^4
\delta^{(4)}(p_1+p_2+p_3-p^{'}_1-p^{'}_2)\,,
\end{equation}
with $| {\cal M}_{ggg \to gg} |^2=| {\cal M}_{gg \to ggg} |^2/d_G$.

Interactions are simulated by the stochastic 
method \cite{XG05,DB91,L93,C02,FCTG08}
according to the interaction (i.e., transition) probabilities
\begin{eqnarray}
\label{p22}
P_{gg\to gg} &=& v_{\rm rel} \frac{\sigma_{gg\to gg}}{N_{\rm test}} 
\frac{\Delta t}{\Delta^3 x} \,, \\
\label{p23}
P_{gg\to ggg} &=& v_{\rm rel} \frac{\sigma_{gg\to ggg}}{N_{\rm test}} 
\frac{\Delta t}{\Delta^3 x} \,,\\
\label{p32}
P_{ggg\to gg} &=& \frac{1}{8 E_1 E_2 E_3} \frac{I_{ggg\to gg}}{N_{\rm test}^2}
\frac{\Delta t}{(\Delta^3 x)^2} \,,
\end{eqnarray}
for $gg\to gg$, $gg\to ggg$, and $ggg\to gg$ processes, respectively.
$v_{\rm rel}=s/(2E_1E_2)$ is the relative velocity between incoming gluons.
$\Delta^3 x$ denotes the volume of the cell element and $\Delta t$
the time step. The probabilities are calculated for every doublet and
triplet in each spatial cell and are compared with random numbers between 
$0$ and $1$ to determine whether or not interactions occur. 
If the random number is smaller than the probability for an interaction,
the interaction occurs and new momenta of outgoing gluons will be sampled
according to Eqs. (\ref{cs22}) or (\ref{m23}). Details of the samplings can be
found in Ref. \cite{XG05}. If the random number is larger than
the probability for an interaction, no interaction will occur.
Particles that do not participate in interactions within $\Delta t$
will propagate as a free streaming. These operations run over all
the cells and time steps until the freezeout condition is reached.

To reduce numerical artifacts, the cell length is set to be smaller than 
the gluon mean free path. For instance, in noncentral Au+Au collisions
with an impact parameter of $b=8.6$ fm, the transverse cell length is 
set to be a constant small value of $\Delta x=\Delta y=0.125$ fm.
The setup of the longitudinal cell length is refreshed before 
each new time step so that there are almost the same number of particles in 
each $\Delta z$ bin with $\Delta z= t[\tanh(\eta_{s2})-\tanh(\eta_{s1})]$,
where $t$ is the time in the center-of-mass frame of a Au+Au collision
and $\eta_s=\frac{1}{2}\ln[(t+z)/(t-z)]$
denotes the space time rapidity. It turns out that the longitudinal cells
are almost equidistant in $\eta_s$ \cite{XG07}, i.e.,
$|\eta_{s2}-\eta_{s1}|\approx {\rm const}$, which implies a nearly
Bjorken-type space time evolution of the gluon matter.

The accurate solution of the Boltzmann equation using the stochastic
method according to Eqs. (\ref{p22})$-$(\ref{p32}) is guaranteed 
if there are enough particles in each cell, because this condition 
is assumed when deriving the interaction probabilities 
Eqs. (\ref{p22})$-$(\ref{p32}) \cite{XG05}.
Because the cell volume has to be small, the real gluon number in each
cell is too small to fulfill the condition for using the stochastic
method. We adopt the test particle method to amplify the particle number
with a factor of $N_{\rm test}$. To leave the physical scale of the gluon
mean free path unchanged, the interaction probabilities are accordingly
reduced by $N_{\rm test}$ and $N_{\rm test}^2$, respectively [see
Eqs. (\ref{p22})$-$(\ref{p32})]. One ``event'' means $N_{\rm test}$
events with parallel propagations. Particles from different events can
interact but with the reduced probabilities, so that each particle still
behaves as a true physical one. On the other hand, when calculating
the real particle number and energy density, the values must be divided 
by $N_{\rm test}$. We set $N_{\rm test}=2400$ for $b=8.6$ fm,
which leads to $|\eta_{s2}-\eta_{s1}|\approx 0.1$ for the longitudinal
cell length.

In the present pQCD simulations the interactions of the gluons are
stopped when the local energy density drops below $e_c$. The value for 
$e_c$ is assumed to be the critical value for the occurrence of 
hadronization, below which parton dynamics is not valid. Because 
a realistic hadronization and a hadronic cascade are not yet included 
in BAMPS, a gluon, which ceases to interact, propagates freely and is 
regarded as a free pion according to the parton-hadron duality picture
($\delta$-function fragmentation of the gluon). Thus, $e_c$ determines
the freezeout condition. The implementation of a Cooper-Frye prescription
for hadronization \cite{P08} and the subsequent UrQMD \cite{urqmd} 
hadronic cascade after the QGP evolution is in progress.

Because the implementation of Bose enhancement into transport calculations
is still technically difficult, gluons are treated as Boltzmann particles
for simplicity. If the gluon system is in thermal equilibrium,
the energy density is $e=48T^4/\pi^2$.
In Ref. \cite{XGS08} we have chosen $e_c=1$ $\rm{GeV\ fm}^{-3}$, which leads
to a critical temperature of $T_c=200$ MeV. To see the possible dependence
of the elliptic flow on the freezeout density we choose, in this article,
an additional moderately lower value of $e_c=0.6$ $\rm{GeV\ fm}^{-3}$
that gives $T_c=175$ MeV.

The initial gluon distribution for BAMPS calculations is the same as 
that chosen in Ref. \cite{XGS08}: an ensemble of minijets \cite{EKL89}
with transverse momenta greater than $p_0=1.4$ GeV, produced via
semihard nucleon-nucleon collisions. The value of $p_0$ is chosen by
matching the final transverse energy per rapidity from the parton cascade
calculations to the experimental data \cite{XGS08}.
We use Glauber geometry with a Woods-Saxon profile and assume independent
binary nucleon-nucleon collisions. The inclusion of quarks in BAMPS is 
straightforward and is in progress. We do not expect large changes in 
the integrated elliptic flow $v_2$, because the quark amount 
at RHIC is initially about $20\%$ only. As quarks interact weaker than 
gluons, we expect that $v_2$ will be slightly smaller than that obtained 
in the present pure gluon matter approach. More discussion about this 
issue appears at the end of the next section.

\section{Elliptic flow and discussions on hadronization}
\label{sec2}
\subsection{Results}
The BAMPS calculations for Au+Au collisions at the RHIC energy
$\sqrt{s_{NN}}=200$ GeV are carried out for a set of discrete impact 
parameters from $b=2$ to $11$ fm (see Appendix \ref{app1}).
The corresponding number of participating nucleons $N_{\rm part}(b)$
is calculated within the wounded nucleon model \cite{BBC76}.
The details are presented in Appendix \ref{app1}. The results from 
the BAMPS calculations with the given impact parameters are used to 
obtain the elliptic flow as a function of the transverse momentum $v_2(p_T)$
and the transverse momentum spectra in different centrality classes.
Calculations with randomly sampled impact parameters are not appropriate
for BAMPS, because a single event is already extremely time consuming 
for large test particle numbers $N_{\rm test}$.

\begin{figure}[ht]
\centerline{\epsfysize=7cm \epsfbox{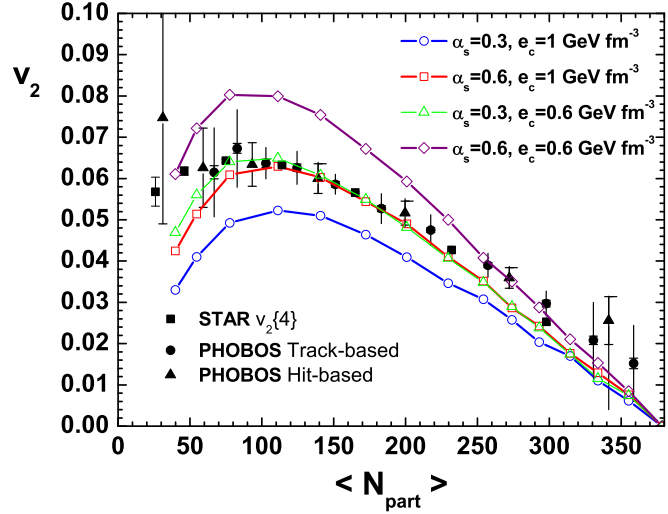}}
\caption{(Color online) Elliptic flow vs $N_{\rm part}$ for Au+Au collisions
at $\sqrt{s_{NN}}=200$ GeV. The points are STAR \cite{star1} and PHOBOS 
\cite{phobos1} data for charged hadrons within $|\eta| < 0.5$ and $|\eta| < 1$,
respectively, whereas the curves with symbols are results for gluons within 
$|\eta| < 1$, obtained from the BAMPS calculations with 
$\alpha_s=0.3$ and $0.6$ and with two freezeout energy densities,
$e_c=0.6$ and $1$ $\rm{GeV\ fm}^{-3}$.
}
\label{v2}
\end{figure}

Figure \ref{v2} shows the elliptic flow 
$v_2=\langle (p_x^2-p_y^2)/p_T^2 \rangle$ as a function of the number
of participating nucleons, $N_{\rm part}$.
The points are STAR \cite{star1} and PHOBOS \cite{phobos1} data for 
charged hadrons within the pseudorapidity intervals $|\eta| < 0.5$ and
$|\eta| < 1$, respectively. The symbols, which are connected with colored
straight lines, are results for gluons within $|\eta| < 1$ ($\eta$ being
identical to the momentum rapidity $y$ for massless gluons) from the BAMPS
calculations for two values of the coupling constant $\alpha_s$ and for two
values of the critical energy density $e_c$ for the freezeout. Especially,
the blue curve with open circles and the red curve with open squares are
the results for ($\alpha_s=0.3$, $e_c=1$ ${\rm GeV\ fm}^{-3}$) and 
($\alpha_s=0.6$, $e_c=1$ ${\rm GeV\ fm}^{-3}$), respectively. These are
exactly the same as shown in Fig. 2 of Ref. \cite{XGS08}. The new results
for $e_c=0.6$ ${\rm GeV\ fm}^{-3}$ will be discussed in the next section.

\begin{figure}[ht]
\centerline{\epsfysize=7cm \epsfbox{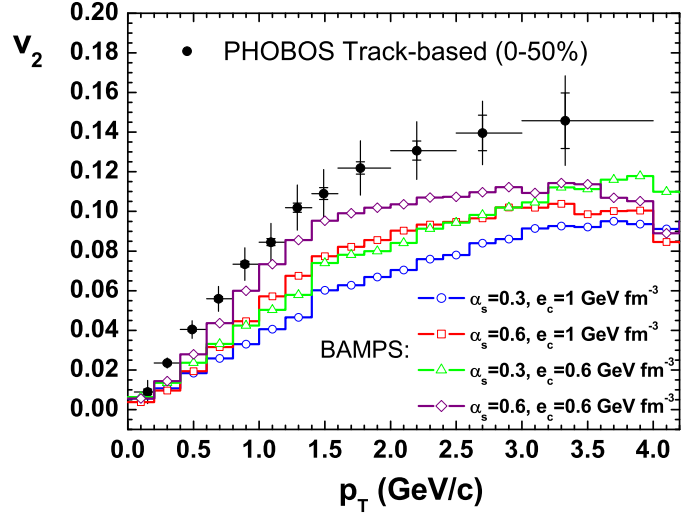}}
\caption{(Color online) Elliptic flow vs transverse momentum for the most
central $50\%$ of Au+Au collisions at $\sqrt{s_{NN}}=200$ GeV.
The points are PHOBOS data \cite{phobos1} for charged hadrons 
within $0 < \eta < 1.5$, whereas the curves with symbols are results 
for gluons within $|\eta| < 1.5$, obtained from the BAMPS calculations with 
$\alpha_s=0.3$ and $0.6$ and with $e_c=0.6$ and $1$ $\rm{GeV\ fm}^{-3}$.
}
\label{v2pt}
\end{figure}

\begin{figure}[ht]
\centerline{\epsfysize=11cm \epsfbox{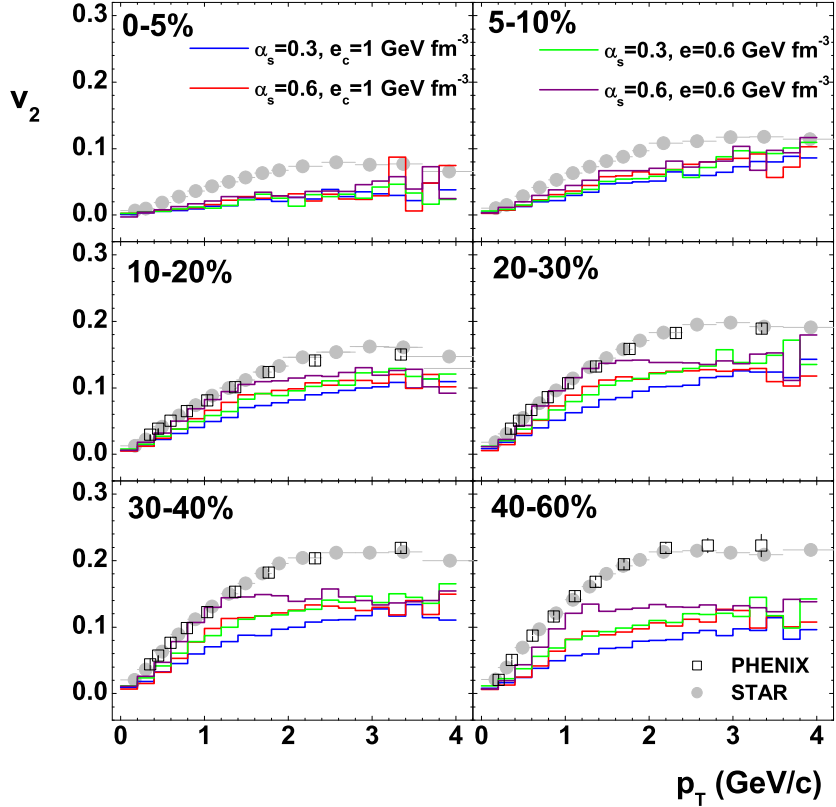}}
\caption{(Color online) Elliptic flow vs transverse momentum for 
different centrality bins of Au+Au collisions at $\sqrt{s_{NN}}=200$ GeV.
The points are PHENIX \cite{PHEN04,phenix} and STAR data \cite{star1} 
for charged
hadrons within $|\eta| < 0.35$ and $|\eta| < 0.5$, whereas the curves are 
results for gluons within $|\eta| < 0.5$, obtained from the BAMPS 
calculations with $\alpha_s=0.3$ and $0.6$ and with 
$e_c=0.6$ and $1$ $\rm{GeV\ fm}^{-3}$.
}
\label{v2pt2}
\end{figure}

Figure \ref{v2pt} shows the elliptic flow as a function of the 
transverse momentum $v_2(p_T)$ for the most
central $50\%$ of Au+Au collisions at $\sqrt{s_{NN}}=200$ GeV.
The points are PHOBOS data \cite{phobos1} for charged hadrons 
within $0 < \eta < 1.5$, whereas the curves with symbols are results 
for gluons within $|\eta| < 1.5$ from BAMPS calculations.
Figure \ref{v2pt2} shows $v_2(p_T)$ in different centrality classes for
particles within $|\eta| < 0.5$.
The PHENIX \cite{PHEN04,phenix} and STAR data \cite{star1} are depicted for 
comparisons with the results from the BAMPS calculations. The latter 
are plotted until $p_T=4$ GeV, because the statistical errors at 
higher $p_T$ are too large.

\begin{figure}[ht]
\centerline{\epsfysize=8cm \epsfbox{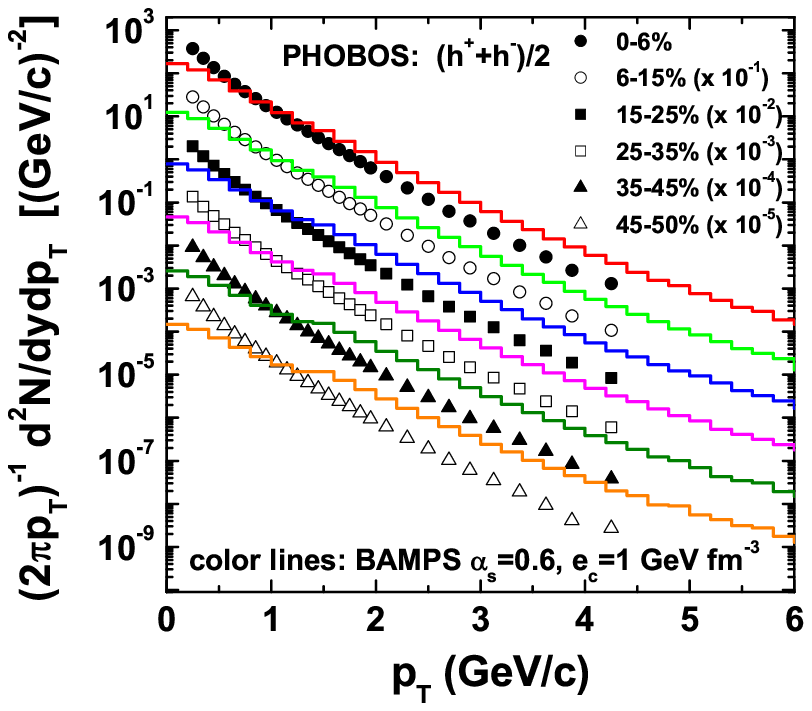}}
\caption{(Color online) Transverse momentum spectra for different centrality
bins of Au+Au collisions at $\sqrt{s_{NN}}=200$ GeV.
The points are PHOBOS data \cite{phobos2} for charged hadrons within 
the rapidity interval $0.2 < y_{\pi^\pm} < 1.4$, whereas the curves are 
results for gluons within $|y| < 1.5$, obtained from the BAMPS 
calculations with $\alpha_s=0.6$ and $e_c=1$ $\rm{GeV\ fm}^{-3}$. 
To compare with the experimental data, the gluon number is reduced 
by a factor of $3$.
}
\label{dndpt}
\end{figure}

Although the integrated $v_2$ of gluons from the BAMPS calculations with
$\alpha_s=0.6$ and $e_c=1$ ${\rm GeV\ fm}^{-3}$ agree perfectly with
the experimental data (see the red curve with open squares in Fig. \ref{v2}),
the differential $v_2(p_T)$ (the red curve with open squares in 
Fig. \ref{v2pt} and red curves in Fig. \ref{v2pt2}) are 
20$-$50$\%$ smaller than the data, especially at 
intermediate momenta $1.5$ GeV $< p_T < 4$ GeV. Correspondingly, 
the $p_T$ spectra of gluons, shown in Fig. \ref{dndpt}, are flatter than
those of hadrons at RHIC.
The larger the centrality, the larger is the difference between the data
and the results from the calculations. 

\begin{figure}[ht]
\begin{center}
\centerline{\epsfysize=7cm \epsfbox{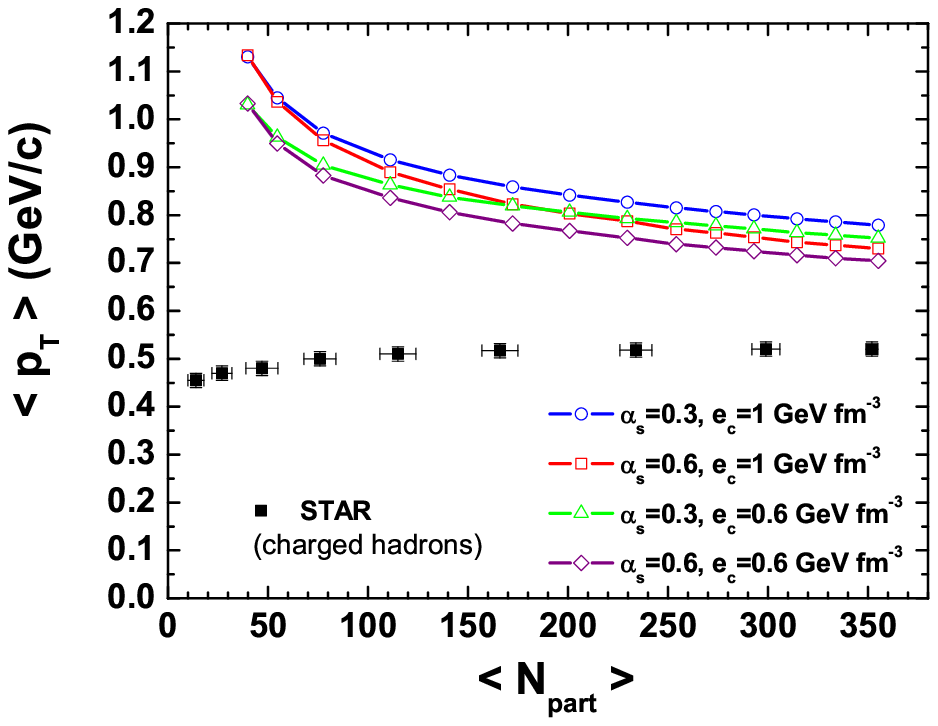}}
\vfill
\centerline{\epsfysize=7cm \epsfbox{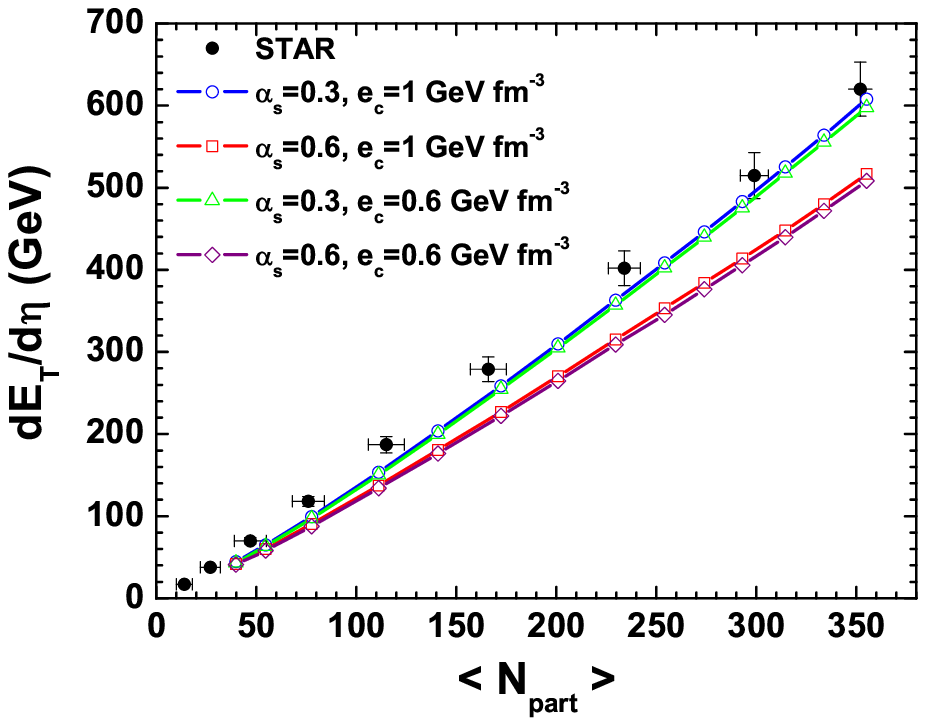}}
\end{center}
\vspace{-0.8cm}
\caption{(Color online) Mean transverse momentum vs $N_{\rm part}$ (upper
panel) and transverse energy per pseudorapidity vs $N_{\rm part}$ (lower
panel) in Au+Au collisions at $\sqrt{s_{NN}}=200$ GeV. 
The points are STAR data \cite{star2} within
$0 < \eta < 1$, whereas the curves with symbols are results for gluons 
within $|\eta| < 1$, obtained from the BAMPS calculations with $\alpha_s=0.3$
and $0.6$ and with $e_c=0.6$ and $1$ $\rm{GeV\ fm}^{-3}$. 
}
\label{meanpt}
\end{figure}

Discrepancy with the RHIC data is also seen in the mean transverse momentum,
$\langle p_T \rangle$, which is shown in the upper panel of 
Fig. \ref{meanpt} as a function of $N_{\rm part}$.
The points are STAR data \cite{star2} within 
$0 < \eta < 1$, whereas the curves with symbols are the BAMPS results 
for gluons within $|\eta| < 1$. The mean $p_T$ of gluons is larger than that
of charged hadrons. On the other hand, the total transverse energy of 
gluons per rapidity at midrapidity, shown in the lower panel of 
Fig. \ref{meanpt}, agrees with the experimental data including hadronic and
electromagnetic components \cite{star2}. We notice that the results 
of $dE_T/d\eta$ with a larger coupling constant of $\alpha_s=0.6$ are 
7$-$15$\%$ smaller than those with $\alpha_s=0.3$ (also smaller than the data)
because of the larger longitudinal work during a less viscous hydrodynamical 
expansion. However, if quarks are included in BAMPS calculations, 
the transverse energy will be slightly enlarged.

\subsection{Discussion}
$dE_T/d\eta$ agrees rather well with the data but $\langle p_T \rangle$ is
larger than the data. This thus indicates that the gluon number per rapidity
is smaller than that of the hadron number per rapidity. The parton-hadron 
duality with one gluon to one pion seems too simple to describe 
the complex hadronization. To obtain $\langle p_T \rangle$ that agrees 
with the data, each gluon is expected to ``fragment'' to 1.5$-$2 pions 
on average. The larger the centrality (peripheral collisions), the more 
fragments are needed, because initial hard gluon jets are less quenched 
in peripheral collisions because of their small reaction size.

Possible fragmentation of each gluon to more pions suggests 
$v_2^{\rm{pion}}(p_T)\approx v_2^{\rm{gluon}}(n p_T)$ and
$dN^{\rm{pion}}(p_T)\sim dN^{\rm{gluon}}(n p_T)$ with $n=$ 1.5$-$2.
The obtained $v_2^{\rm{pion}}(p_T)$ for low $p_T < 1$ GeV would come closer
to the experimental data. Because the integrated $v_2$ is determined
mostly by the lower $p_T$ particles, the integrated $v_2$ of pions will 
be nearly the same as that of gluons. On the other hand, the 
$v_2^{\rm{pion}}(p_T)$ for $p_T > 1$ GeV would be still 
20$-$50$\%$ smaller than the data. The discrepancy is larger for more
peripheral collisions (see Fig. \ref{v2pt2}). It seems that quark 
recombination models \cite{K0203,VM03,F03,HY03} for hadronization are
appropriate for intermediate $p_T$ gluons, if one assumes that each gluon
must first be converted to a quark-antiquark pair $g\to q\bar q$ via 
$v_2^{\rm{quark}}(p_T/2)\approx v_2^{\rm{gluon}}(p_T)$. Then the quark
recombination models would give 
$v_2^{\rm{pion}}(p_T)\approx 2 v_2^{\rm{quark}}(p_T/2)) 
\approx 2 v_2^{\rm{gluon}}(p_T)$.

We also must stress that the results discussed above are obtained from
the present calculations without quark degrees of freedom.
The inclusion of quarks will certainly change the freezeout temperature,
because the freezeout energy density is proportional to the degeneracy
factor of partons, $e_c \sim (d_G+d_Q) T_c^4$, where $d_G=16$ for gluons
and $d_Q=24$ for quarks with two flavors. For the same freezeout energy
density the freezeout temperature will be a factor of $1.26$ smaller when
the quarks are included and if all the partons are chemically equilibrated.
Although the initial energy is slightly enlarged when including quarks,
we do not expect much stronger transverse flow at freezeout, as long as
the shear viscosity of the parton system is the same with or without quarks.
Therefore, the inclusion of quarks might reduce the mean $p_T$ by a factor
of $1.26$, which then would lead to a fair agreement between the calculated
and the experimentally measured mean $p_T$ via parton-hadron duality.
If the integrated $v_2$ is still comparable with the experimental data,
then $v_2(p_T)$ will also come to a closer agreement with the data.
Work in this direction is in progress.

The parton-hadron duality picture is consistent with the hadronization in
a thermalized expanding matter, which assumes that both constituents of
the bulk matter at the beginning and at the end of the hadronization will
be distributed according to thermal statistics. If the temperature
does not change much during the phase transition, the integrated $v_2$, 
$v_2(p_T)$ and the $p_T$ spectra of pions are almost the same as those 
of partons. Although the energy density of pions is smaller than that 
of partons, the matter will expand continuously during the hadronization, 
which leaves the total energy, and the total particle number and, thus, the 
entropy of the constituents approximately unchanged. Within the picture 
of a thermalized expanding source the dependence of $v_2(p_T)$ on 
the hadron mass was also understood \cite{H01} and agrees with the RHIC 
data at lower $p_T < 1.5$ GeV \cite{PHEN04,STAR04-1}.

On the other hand, the STAR data \cite{star92,star1} confirmed quark
recombination models \cite{K0203,VM03,F03,HY03} for hadronization of quarks
with intermediate $p_T\sim$ 1$-$2 GeV. These models lead to
$v_2^{\rm{hadron}}(p_T)\approx n v_2^{\rm{quark}}(p_T/n)$ for $n=2$ for 
mesons and for $n=3$ for baryons, which is obviously different from the
picture of a thermalized expanding source. How the transition from
one to another hadronization scenario proceeds is not clear. Also, how gluons
convert into quarks and antiquarks in the recombination models has to be
understood and modeled.

We emphasize that how the partons hadronize is essential to obtain 
a complete picture of the generation of elliptic flow and to understand
the transformation of the $v_2(p_T)$, the $p_T$ spectra, and the mean $p_T$
from the partonic to the hadronic phase in a consistent manner. 
Quantitative calculations modeling hadronization may give a more detailed 
understanding of this subject. As a starting point, an algorithm employing 
a Cooper-Frye prescription \cite{P08} will be implemented in BAMPS 
in the near future. Detailed comparisons with results from recent
calculations using viscous hydrodynamics \cite{DT08} will be given.

Finally, the gluons in the present BAMPS calculations are assumed as
Boltzmann particles rather than bosons because of numerical difficulties.
Bose enhancement increases the gluon density by a factor of
$\zeta(3)\approx 1.202$ for a thermal gluon system. Moreover, the 
enhancement at low $p_T$ is considerably larger than that at large
$p_T$. This may also have a moderate effect on the $p_T$ dependence of
the elliptic flow and thus will be investigated in more detail.

\section{Dependence of the elliptic flow on the freezeout condition}
\label{sec3}
The elliptic flow $v_2$ will be continuously built up if the spatial 
anisotropy of the particle system still exists and sufficiently strong
interactions among particles are present. The buildup of $v_2$ does not
necessarily mean that $v_2$ becomes always larger. After the time when 
the eccentricity crosses zero the system will turn spatially anisotropic
again but with the sign opposite to the eccentricity of the initial one.
Further evolution will then decrease $v_2$. Whether or not a generation
of negative $v_2$ occurs depends on whether the kinetic freezeout of
the particle system happens after or before the time when the initial
eccentricity crosses zero.

In this section we discuss the dependence of the elliptic flow on
the freezeout condition in Au+Au collisions at the RHIC energy.
In the present BAMPS calculations the kinetic and chemical freezeouts
are assumed to occur at the same time when the local energy density
drops below the chosen cutoff value of $e_c$. The cutoff $e_c$ serves as 
a critical energy density, at which gluons are converted to hadrons.
After the phase transition no hadronic interactions take place in 
the present version of BAMPS. We have already speculated that 
a realistic modeling of hadronization may change the pattern of 
$v_2(p_T)$ in various ways. Moreover, a hadronic cascade simulating
the decoupling stage should be included \cite{ZBS05,H06,BBFF07}. 
With the hadronic cascade, the kinetic freezeout happens later than 
the hadronization and the final elliptic flow will likely be different 
from that obtained in the present BAMPS calculations. However, 
this difference is expected to be rather marginal, because the viscosity 
in the hadron gas is much larger than that in the QGP.

Although the freezeout is considered in a simple way in the present BAMPS,
it is useful to study the uncertainty in the calculated $v_2$,
if the value of the critical energy density $e_c$ is varied to change 
the lifetime of the QGP. If the freezeout condition constrains the 
elliptic flow, then it also constrains the to be deduced shear viscosity 
of the QGP. For this purpose, we have performed calculations with 
$e_c=0.6$ ${\rm GeV\ fm}^{-3}$ in addition to those with 
$e_c=1$ ${\rm GeV\ fm}^{-3}$. The results are shown in 
Figs. \ref{v2}$-$\ref{v2pt2} and \ref{meanpt} by green curves with open
triangles ($\alpha_s=0.3$) and purple curves with open diamonds
($\alpha_s=0.6$).

\begin{figure}[ht]
\centerline{\epsfysize=7cm \epsfbox{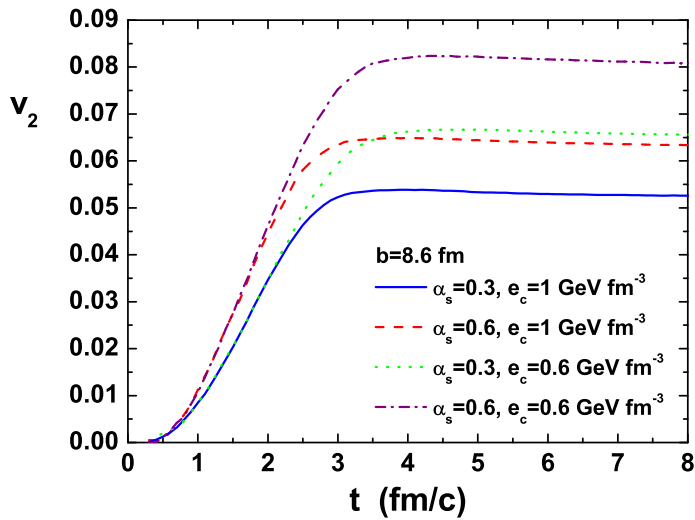}}
\caption{(Color online) Time evolution of the elliptic flow from the BAMPS
calculations in Au+Au collisions at $\sqrt{s_{NN}}=200$ GeV at 
an impact parameter of $b=8.6$ fm.
}
\label{v2t}
\end{figure}

We realize that the $v_2$ results with $\alpha_s=0.3$ and 
$e_c=0.6$ ${\rm GeV\ fm}^{-3}$ (green curves with open triangles) are 
almost identical to those with $\alpha_s=0.6$ and 
$e_c=1$ ${\rm GeV\ fm}^{-3}$ (red curves with open squares).
Stronger interactions or longer QGP phase leads to the same final values
of $v_2$. Figure \ref{v2t} shows the $v_2$ generation as a function of time
in a Au+Au collision with an impact parameter of $b=8.6$ fm.
No decrease of $v_2$ is observed. This indicates that the freezeout occurs
before the initial spatial anisotropy vanishes. The saturation of $v_2$
begins at $2.5$ fm/c for $e_c=1$ ${\rm GeV\ fm}^{-3}$ and later at $3$ fm/c
for $e_c=0.6$ ${\rm GeV\ fm}^{-3}$. The continuous increase of $v_2$ after
$2.5$ fm/c in the case for $e_c=0.6$ ${\rm GeV\ fm}^{-3}$ is as strong as
that before. The difference of the pressure gradient between the $x$ and $y$
directions is still large at $2.5$ fm/c, whereas at this time the freezeout at
$e_c=1$ ${\rm GeV\ fm}^{-3}$ is nearly complete. Therefore, the uncertainty
in the final elliptic flow due to the different freezeout condition is
not small. In addition to the hadronization, the time scale when
the hadronization occurs affects also the absolute value of $v_2$ as well as
the transverse momentum dependence of $v_2$.

A detailed study of the elliptic flow is important, because this collective
effect of QCD matter quantitatively constrains the shear viscosity of
the medium. In Ref. \cite{XGS08} we have demonstrated within the BAMPS
calculations that the shear viscosity to the entropy density ratio
$\eta/s$ is nearly constant in time and approximately depends only on
the coupling constant $\alpha_s$: $\eta/s \approx 0.15$ for $\alpha_s=0.3$
and $\eta/s \approx 0.08$ for $\alpha_s=0.6$. From Fig. \ref{v2} we see
that with either $\alpha_s=0.3$ and $e_c=0.6$ ${\rm GeV\ fm}^{-3}$ or
$\alpha_s=0.6$ and $e_c=1$ ${\rm GeV\ fm}^{-3}$ the same values of the final
$v_2$ are obtained and which agree very well with the experimental data.
Therefore, according to the present study, $\eta/s$ is most probably lying
between $0.15$ and $0.08$.

If the freezeout occurs at a lower energy density (or temperature), one
would expect that the transverse momentum spectra will become softer
and the final transverse energy will be smaller. From the lower panel
of Fig. \ref{meanpt} we see that $dE_T/d\eta$ for 
$e_c=0.6$ ${\rm GeV\ fm}^{-3}$ are only slightly smaller than 
those for $e_c=1$ ${\rm GeV\ fm}^{-3}$.
The differences in the transverse momentum spectra (not shown) are also tiny.
The further decrease 
of the local energy density from $1$ to $0.6$ ${\rm GeV\ fm}^{-3}$
due to the longitudinal work done by the pressure is marginal, because
at the late stage of expansion the system becomes dilute and thus the
work done is small. Free streaming increases the transverse flow
and thus effectively decreases the local energy density.
Therefore, a lower energy density cutoff for the freezeout does not lead to
a much smaller total transverse energy and mean transverse momentum, as 
seen in Fig. \ref{meanpt}. On the other hand, the $v_2(p_T)$ for $\alpha_s=0.6$
and $e_c=0.6$ ${\rm GeV\ fm}^{-3}$, shown by the purple curve with open
diamonds in Fig. \ref{v2pt} and by purple curves in Fig. \ref{v2pt2}, 
agree well with the data at the low $p_T < 1.5$ GeV, whereas the 
integrated $v_2$ (see the purple curve with open diamonds in 
Fig. \ref{v2}) overestimate the data.

We note that within the same description for the gluon interactions 
the energy loss of high $p_T$ gluons is found to be in good agreement
with the results from the GLV formalism when $\alpha_s=0.3$ is 
used \cite{FXG08}. Hence both jet-quenching phenomena and the buildup of
elliptic flow are rather well described in BAMPS parton cascade calculations.

\section{Summary}
\label{sum}
Employing the pQCD based parton cascade BAMPS that includes pQCD
bremsstrahlung and its backreaction we have calculated the elliptic flow
$v_2$ in Au+Au collisions at $\sqrt{s_{NN}}=200$ GeV. Hadronization is 
assumed to happen at a critical energy density, $e_c$. A gluon is then
fragmented into a pion according to the parton-hadron duality assumption.
The hadron $v_2$ is therefore identical to the final gluon $v_2$ within the
present BAMPS calculations. 

We found that 
whereas the final gluon $v_2$ from the calculations with $\alpha_s=0.6$ and
$e_c=1$ ${\rm GeV\ fm}^{-3}$ agree well with the data, the transverse 
momentum dependence $v_2(p_T)$ are 20$-$50$\%$ lower then the data in each 
centrality class. The final gluon transverse momentum spectra are harder
than the data and also the final gluon mean transverse momenta are 40$-$100$\%$
larger than the data. However, the final gluon transverse energy per 
rapidity at midrapidity agrees with the data. This indicates that the
hadronization process via the parton-hadron duality is not justified. 
A realistic hadronization that transforms a gluon to 1.5$-$2
pions on average would be expected for low $p_T$ gluons, whereas at
intermediate $p_T$, quark recombination would be a better scenario of 
hadronization. In addition, the inclusion of quark degrees 
of freedom would reduce the final parton mean $p_T$ and would lead to 
a better agreement between the calculated and measured $v_2(p_T)$.
This must be demonstrated in a new version of BAMPS calculations
including the quark dynamics and employing various hadronization scenarios.

The value of the final gluon $v_2$ depends on the freezeout condition,
i.e., $e_c$. The pressure gradient difference in the $x$ and $y$ directions
at $e_c=1$ ${\rm GeV\ fm}^{-3}$ is still large enough to further increase
the elliptic flow if $e_c$ is changed to a smaller but still reasonable 
value. We observed that
the final gluon $v_2$ and $v_2(p_T)$ are almost the same in the calculations
with either $\alpha_s=0.6$ and $e_c=1$ ${\rm GeV\ fm}^{-3}$ or
$\alpha_s=0.3$ and $e_c=0.6$ ${\rm GeV\ fm}^{-3}$. Stronger interactions
or a later freezeout leads to the same elliptic flow. This outlines
the uncertainty in the extraction of the shear viscosity in the QGP:
The shear viscosity to the entropy density ratio will most probably be
between $0.08$ and $0.15$.

We furthermore conclude that adding quark degrees of freedom into 
the dynamical evolution of the QCD matter with a detailed understanding of
the hadronization of gluons and quarks will be helpful in explaining 
the viscous facets of the final hadron elliptic flow and in extracting 
the shear viscosity to the entropy density ratio of the QGP. 
The results presented in this article motivate more detailed 
investigations of this issue in future works.

\acknowledgments
The authors thank M. Bleicher, G. Burau, and H. St\"ocker for enlightening 
discussions.
The BAMPS simulations were performed at the Center for Scientific 
Computing of Goethe University.
This work was financially supported by the Helmholtz International Center
for FAIR within the framework of the LOEWE program (Landes-Offensive zur
Entwicklung Wissenschaftlich-\"okonomischer Exzellenz) launched
by the State of Hesse.

\appendix

\section{Number of participating nucleons $N_{\rm part}$ and 
centrality classes}
\label{app1}
The number of participating nucleons $N_{\rm part}(b)$ in an A+B
collision at an impact parameter of $b$ is calculated within the description
of wounded nucleons \cite{BBC76}:
\begin{equation}
N_{\rm part}(b)= \int d^2 s \, n_{\rm part}(\vec{s}, \vec{b})\,,
\end{equation}
where
\begin{equation}
n_{\rm part}(\vec{s}, \vec{b}) = T_A(\vec{s}) \,
\left [ 1- e^{- \sigma_H \, T_B(\vec{b}-\vec{s})} \right ]+ 
T_B(\vec{b}-\vec{s}) \,
\left [ 1- e^{- \sigma_H \, T_A(\vec{s})} \right ]\,.
\end{equation}
$\sigma_H$ denotes the nucleon-nucleon total inelastic cross section with
diffraction production excluded and is set to be $\sigma_H=42$ mb.
$T_A(\vec{s})$, also $T_B(\vec{s})$, is the thickness function defined as
\begin{equation}
T_A(\vec{s})=\int_{-\infty}^{\infty} dz\, \rho_A(\vec{s}, z)\,,
\end{equation}
where $\rho_A(\vec{s}, z)$ is the single nucleon density and
$\int d^3r \, \rho_A(\vec{r})=\int d^2s \, T_A(\vec{s})=A$. 
We use the Woods-Saxon function for the nucleon density
\begin{equation}
\label{wsf}
\rho_A(\vec{r})=\rho_A(r)=\frac{n_0}{1+e^{\frac{r-R_A}{d}}}\,,
\end{equation}
where $d=0.54$ fm,  $n_0=0.17$ ${\rm fm}^{-3}$, and 
$R_A=1.12 A^{1/3}-0.86 A^{-1/3}=6.37$ fm for $A=197$ of a Au nucleus.

We use $N_{\rm part}(b)$ to make a relation between the intervals of the
centrality class and the impact parameter $b$. For an interval of $b$,
$[\bar{b}_1 ; \bar{b}_2]$, the average $N_{\rm part}$ and the average 
impact parameter can be calculated by
\begin{eqnarray}
\label{av_npart}
&& \langle N_{\rm part} \rangle |_{[\bar{b}_1 ; \bar{b}_2]}=
\frac{\int_{\bar{b}_1}^{\bar{b}_2} db \, b N_{\rm part}(b)}
{\int_{\bar{b}_1}^{\bar{b}_2} db\, b}=\frac{2}{\bar{b}_2^2-\bar{b}_1^2}\,
\int_{\bar{b}_1}^{\bar{b}_2} db \, b N_{\rm part}(b) \\
\label{av_b}
&& \langle b \rangle |_{[\bar{b}_1 ; \bar{b}_2]}=
\frac{\int_{\bar{b}_1}^{\bar{b}_2} db \, b^2}
{\int_{\bar{b}_1}^{\bar{b}_2} db\, b}=
\frac{2}{3} \, \frac{\bar{b}_2^3-\bar{b}_1^3}{\bar{b}_2^2-\bar{b}_1^2} \,.
\end{eqnarray}
The number of events, $N_{\rm events}$, within $[\bar{b}_1 ; \bar{b}_2]$ is
proportional to $\bar{b}_2^2-\bar{b}_1^2$. Therefore,
\begin{equation}
\label{av_event}
\frac{N_{\rm events}|_{[\bar{b}_1 ; \bar{b}_2]}}{N_{\rm events}^M}
=\frac{\bar{b}_2^2-\bar{b}_1^2}{{\rm Max}\{\bar{b}_{j+1}^2-\bar{b}_j^2\}}\,,
\end{equation}
where $N_{\rm events}^M$ is the maximum of $N_{\rm events}$ within
all the intervals $[\bar{b}_j ; \bar{b}_{j+1}]$, $j=1,2, \cdots$.

We obtain the intervals of $b$ corresponding with each centrality class by
tuning $\bar{b}_j$s, so that 
$\langle N_{\rm part} \rangle |_{[\bar{b}_j ; \bar{b}_{j+1}]}$,
$\langle b \rangle |_{[\bar{b}_j ; \bar{b}_{j+1}]}$ and 
$N_{\rm events}|_{[\bar{b}_j ; \bar{b}_{j+1}]}/N_{\rm events}^M$ from
Eqs. (\ref{av_npart})$-$(\ref{av_event}) are comparable with the experimental 
data given for each centrality classes. Tables \ref{table1} and \ref{table2}
show the correspondences of the intervals of $b$ with the centrality classes
and the comparisons between the calculated values and the data from
PHOBOS \cite{phobos2} and STAR \cite{star1}.
\begin{table}[htbp]
\begin{center}
\caption{
\label{table1}
Correspondence of the impact parameter intervals with the centrality classes
by matching the PHOBOS data \cite{phobos2}.
}
\begin{tabular}{l|c c c c c c}
\hline 
\hline 
Centrality & 0$-$6\% & 6$-$15\% & 15$-$25\% & 25$-$35\% & 35$-$45\% & 
45$-$50\% \\
$[\bar{b}_j; \bar{b}_{j+1}]$ (fm) &
[0; 3.16] & [3.16; 5.23] & [5.23; 6.95] & [6.95; 8.31] &
[8.31; 9.46] & [9.46; 9.98] \\
\hline
$\langle  N_{\rm part} \rangle$ PHOBOS \cite{phobos2} &
$344 \pm 11$ & $276 \pm 9$ & $200 \pm 8$ & $138 \pm 6$ & $93 \pm 5$  &
$65 \pm 4$\\
$\langle  N_{\rm part} \rangle$ Eq. (\ref{av_npart}) &
350 & 282 & 208 & 147 & 101 & 74 \\
$N_{\rm events}/N_{\rm events}^M$ PHOBOS \cite{phobos2} &
0.505 & 0.883 & 0.997 & 0.997 & 1.0 & 0.5 \\
$N_{\rm events}/N_{\rm events}^M$ Eq. (\ref{av_event}) &
0.477 & 0.829 & 1.0 & 0.99 & 0.98 & 0.48 \\
\hline
\hline
\end{tabular}
\end{center}
\end{table}
\begin{table}[htbp]
\begin{center}
\caption{
\label{table2}
Correspondence of the impact parameter intervals with the centrality classes
by matching the STAR data \cite{star1}.
}
\begin{tabular}{l|c c c c c c c}
\hline
\hline 
Centrality & 0$-$5\% & 5$-$10\% & 10$-$20\% & 20$-$30\% & 30$-$40\% & 
40$-$50\% & 50$-$60\% \\
$[\bar{b}_j; \bar{b}_{j+1}]$ (fm) &
[0; 3.22] & [3.22; 4.68] & [4.68; 6.59] & [6.59; 8] & [8; 9.36] &
[9.36; 10.42] & [10.42; 11.07] \\
\hline
$\langle  N_{\rm part} \rangle$ STAR \cite{star1} &
$352 \pm 6$ & $298 \pm 10$ & $232 \pm 10$ & $165 \pm 12$ & $114 \pm 12$ &
$75 \pm 11$ & $46 \pm 9$  \\
$\langle  N_{\rm part} \rangle$ Eq. (\ref{av_npart}) &
349 & 293 & 226 & 160 & 108 & 69 & 46 \\
$\langle b \rangle$ (fm) STAR \cite{star1} &
$2.3 \pm 0.2$ & $4.2 \pm 0.3$ & $5.9 \pm 0.3$ & $7.6 \pm 0.4$ &
$9.0 \pm 0.5$ & $10.2 \pm 0.5$ & $11.3 \pm 0.6$ \\
$\langle b \rangle$ (fm) Eq. (\ref{av_b}) &
2.15 & 3.99 & 5.69 & 7.32 & 8.70 & 9.90 & 10.75 \\
\hline
\hline
\end{tabular}
\end{center}
\end{table}

BAMPS calculations are carried out for a set of discrete impact parameters
$b=2$, 2.8, 3.4, 4, 4.5, 5, 5.6,
6.3, 7, 7.8, 8.6, 9.6, 10.4, and 11 fm. The average of an observable 
${\cal O}$ in one centrality class is calculated by the integral
\begin{equation}
\langle {\cal O} \rangle |_{[\bar{b}_1 ; \bar{b}_2]}
=\frac{\int_{\bar{b}_1}^{\bar{b}_2} db \, b {\cal O}(b)}
{\int_{\bar{b}_1}^{\bar{b}_2} db\, b}
\end{equation}
using the trapezoid formula.


\end{document}